\documentclass{PoS}

\usepackage{amsmath}
\usepackage{units}

\newcommand{\VC}[1]{%
  \begin{tabular}[c]{l}%
    #1%
  \end{tabular}
}

\DeclareRobustCommand*\diff[2][]{%
   \mathop{
     \mathrm{d}^{#1}
     \mskip-0.2\thinmuskip
    #2}\nolimits
}

\newcommand\3[1]{\boldsymbol{#1}}

\newcommand{\T}[1]{\boldsymbol{#1}_{\text{T}}}
\newcommand{\Tsc}[1]{#1_{\text{T}}}

\newcommand{\bmax}{b_{\rm max}}
\newcommand\bstar{\3{b}_*}
\newcommand\bstarsc{b_*}
\newcommand\mubstar{\mu_{\bstarsc}}

\title{TMD factorization and evolution at large $b_T$}

\ShortTitle{TMD factorization and evolution at large $b_T$}

\author{\speaker{John COLLINS}\\
        104 Davey Lab., Penn State University, University Park PA
        16802, USA \\
        E-mail: \email{jcc8@psu.edu}}

\author{Ted C. Rogers\\
       Theory Center, Jefferson Lab, 12000 Jefferson Avenue, Newport
       News, VA 23606, USA\\
       Department of Physics, Old Dominion University, Norfolk, VA
       23529, USA \\
      E-mail: \email{trogers@odu.edu}
}

\abstract{%
In using transverse-momentum-dependent (TMD) parton densities and
fragmentation functions, important non-perturbative information is at
large transverse position $b_T$.  This concerns both the TMD functions
and their evolution.  Fits to high energy data tend to predict too
rapid evolution when extrapolated to low energies where larger values
of $b_T$ dominate.  I summarize a new analysis of the issues.  It
results in a proposal for much weaker $b_T$ dependence at large $b_T$
for the evolution kernel, while preserving the accuracy of the
existing fits.  The results are particularly important for using
transverse-spin-dependent functions like the Sivers function.
}

\FullConference{XXIII International Workshop on Deep-Inelastic Scattering\\
		 27 April - May 1 2015\\
		 Dallas, Texas}

\begin{document}

\section{Introduction}

This talk summarized some recent work \cite{Collins:2014jpa} in
collaboration with Ted Rogers.

The overall motivation is to understand the evolution of
transverse-momentum-dependent (TMD) parton densities (etc) especially
at the relatively low values of $Q$ that have considerable current
interest, as can be seen from many talks in this session on spin
physics.  We are particularly concerned with the non-perturbative part
of the evolution, since there appear to be inconsistencies between the
evolution found in fits \cite{Landry:2002ix,Konychev:2005iy} to higher
$Q$ Drell-Yan data, and the slower evolution preferred (e.g.,
\cite{Sun:2013hua,Aidala:2014hva}) by more recent
data at lower $Q$.  When TMD functions
are Fourier transformed into a space of transverse coordinates
$\T{b}$, non-perturbative contributions are at large $\T{b}$.

Our aim was to to try to preserve good fits to the Drell-Yan data,
while also agreeing with the lower energy data and satisfying
non-perturbative constraints from the theory side.

\section{Review of TMD factorization and the organization of
  non-perturbative information}

We use the following TMD factorization formula for the Drell-Yan cross
section differential in the lepton pair momentum $q^\mu$ and lepton
angle:
\begin{equation}
\label{eq:fact}
  \frac{ \diff{\sigma} }{ \diff[4]{q}\diff{\Omega} } 
  =  \frac{2}{s} \sum_j
      \frac{ \diff{\hat{\sigma}_{j\bar{\jmath}}}(Q,\mu\mapsto Q) }{ \diff{\Omega} }
        \int 
        e^{i\T{q}\cdot \T{b} }
        ~ \tilde{f}_{j/A}(x_A,\T{b};Q^2,Q) 
        ~ \tilde{f}_{\bar{\jmath}/B}(x_B,\T{b};Q^2,Q)
        \diff[2]{\T{b}},
\end{equation}
valid when $\Tsc{q} \ll Q$ and polarization effects are ignored.  The
functions $\tilde{f}(x,\T{b},Q^2,Q)$ are the Fourier transformed
parton densities (pdfs) to $\T{b}$, with the CSS $\zeta$ and $\mu$
parameters set to $Q^2$ and $Q$. The perturbative hard scattering
factor is $\diff{\hat{\sigma}}(Q,\mu\mapsto Q)$.  The TMD pdfs obey
an evolution equation of the form
\begin{equation}
\label{eq:evol}
  \frac{ \diff{ \ln \tilde{f}_{f/H}(x,\Tsc{b}; Q^2; Q) } }
       { \diff{ \ln Q } }
  = 
  \gamma(\alpha_s(Q)) + \tilde{K}(\Tsc{b};Q)
  =
  \gamma(\alpha_s(Q))
  - \int_{\mu_b}^Q \frac{\diff{\mu}}{\mu} \gamma_K(\alpha_s(\mu))
  + \tilde{K}(\Tsc{b};\mu_b).
\end{equation}
The strongly universal function $\tilde{K}(\Tsc{b};\mu)$ controls
evolution of the shape of the TMD functions, and its behavior at large
$\Tsc{b}$ is the primary concern of our work.  In the right-most part
of (\ref{eq:evol}), a renormalization-group transformation
to scale $\mu_{\bstarsc} \propto 1/\bstarsc$
was applied
to remove large logarithms.

Non-perturbative information is (a) in the values of the TMD pdfs and
of $\tilde{K}$ at large $\Tsc{b}$, and (b) from ordinary pdfs that
appear in the OPE that gives the TMD pdfs at small $\Tsc{b}$.

To separate non-perturbative contributions to evolution, we use the
CSS method to write:
\begin{equation}
\label{eq:evol.bstar}
  \frac{ \diff{ \ln \tilde{f}_{f/H}(x,\Tsc{b}; Q^2; Q) } }
       { \diff{ \ln Q } }
  =
  \gamma(\alpha_s(Q))
  - \int_{\mu_{\bstarsc}}^Q \frac{\diff{\mu}}{\mu} \gamma_K(\alpha_s(\mu))
  + \tilde{K}(\bstarsc;\mu_{\bstarsc})
  - g_K(\Tsc{b};\bmax),
\end{equation}
where a smooth cutoff on the perturbative part is provided by
\begin{math}
  \bstarsc = { \Tsc{b} } / { \sqrt{ 1 + \Tsc{b}^2/\bmax^2} }.  
\end{math}
Perturbative calculations give the first three terms on the right of
(\ref{eq:evol.bstar}), while fits to data are made for a parameterized
form for $g_K(\Tsc{b};\bmax)$, which includes the non-perturbative
large $\Tsc{b}$ contributions.

The predictive power of this TMD factorization formalism beyond the
calculable perturbative contributions is from two sources.  First is
the universality of pdfs between reactions.  Second is the
\emph{strong} universality of $\tilde{K}$, i.e., its lack of
dependence on the reaction, on the hadron and quark flavor, and on spin
and $x$.  The $Q$ dependence of $\tilde{K}$ is governed by the
perturbative function $\gamma_K$, leaving non-perturbative dependence
as a function of $\Tsc{b}$ only.

Common choices of $\bmax$ are $\unit[0.5]{GeV^{-1}}$ and
$\unit[1.5]{GeV^{-1}}$.  Observe that if $\bmax$ is chosen to be too
conservatively small, then fitting of $g_K$ to data includes
reproducing the full $\tilde{K}(\Tsc{b})$ in a region of $\Tsc{b}$
that is still accessible to perturbative calculations.

\section{Geography of evolution}

\begin{figure}
  \centering
  \newcommand\thisscale{0.3}
  \begin{tabular}{c@{\hspace*{1cm}}c@{\hspace*{1cm}}c}
     $\Tsc{q}$
     &
     \VC{\includegraphics[scale=\thisscale]{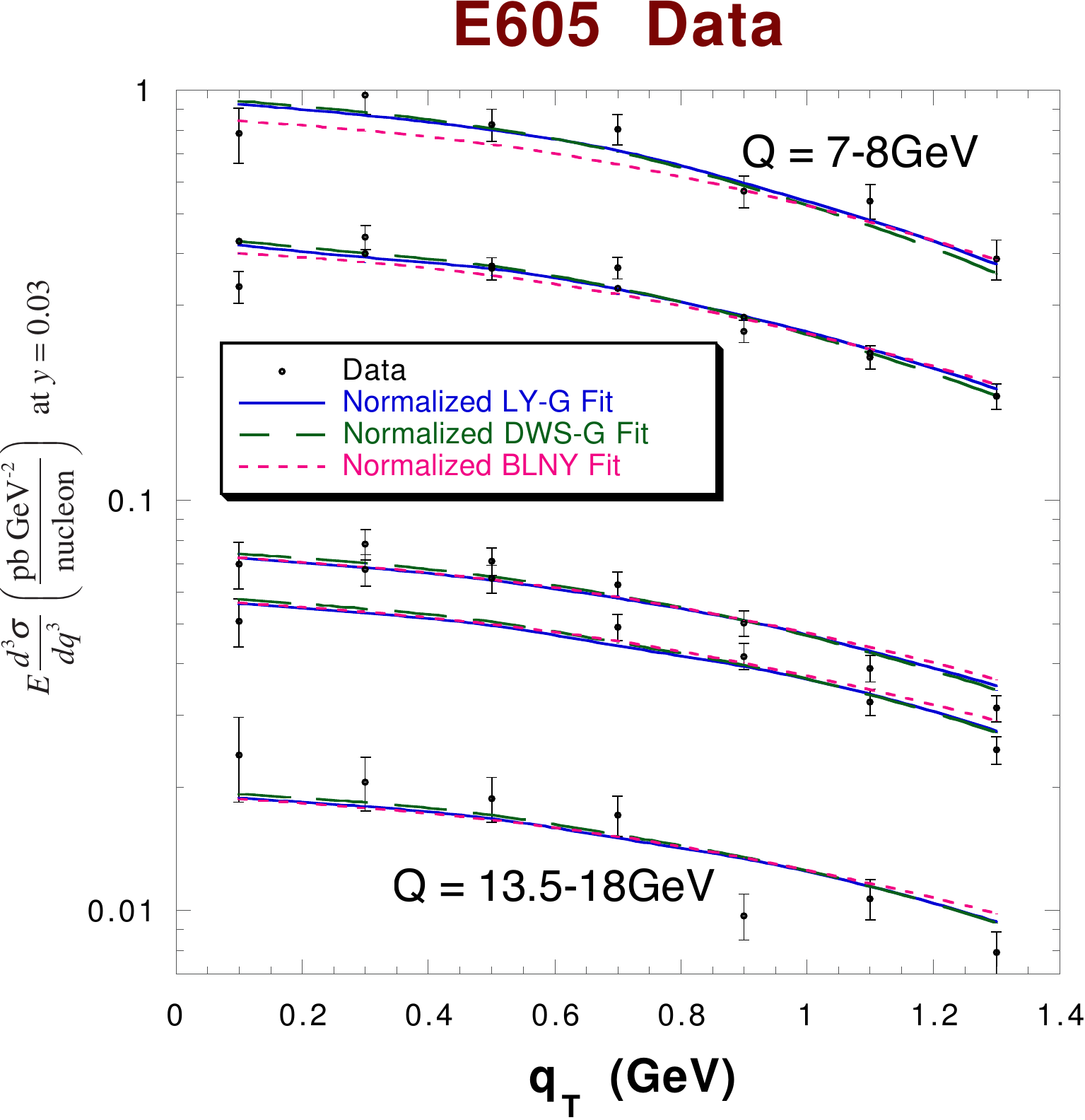}}
     &
     \VC{\includegraphics[scale=\thisscale]{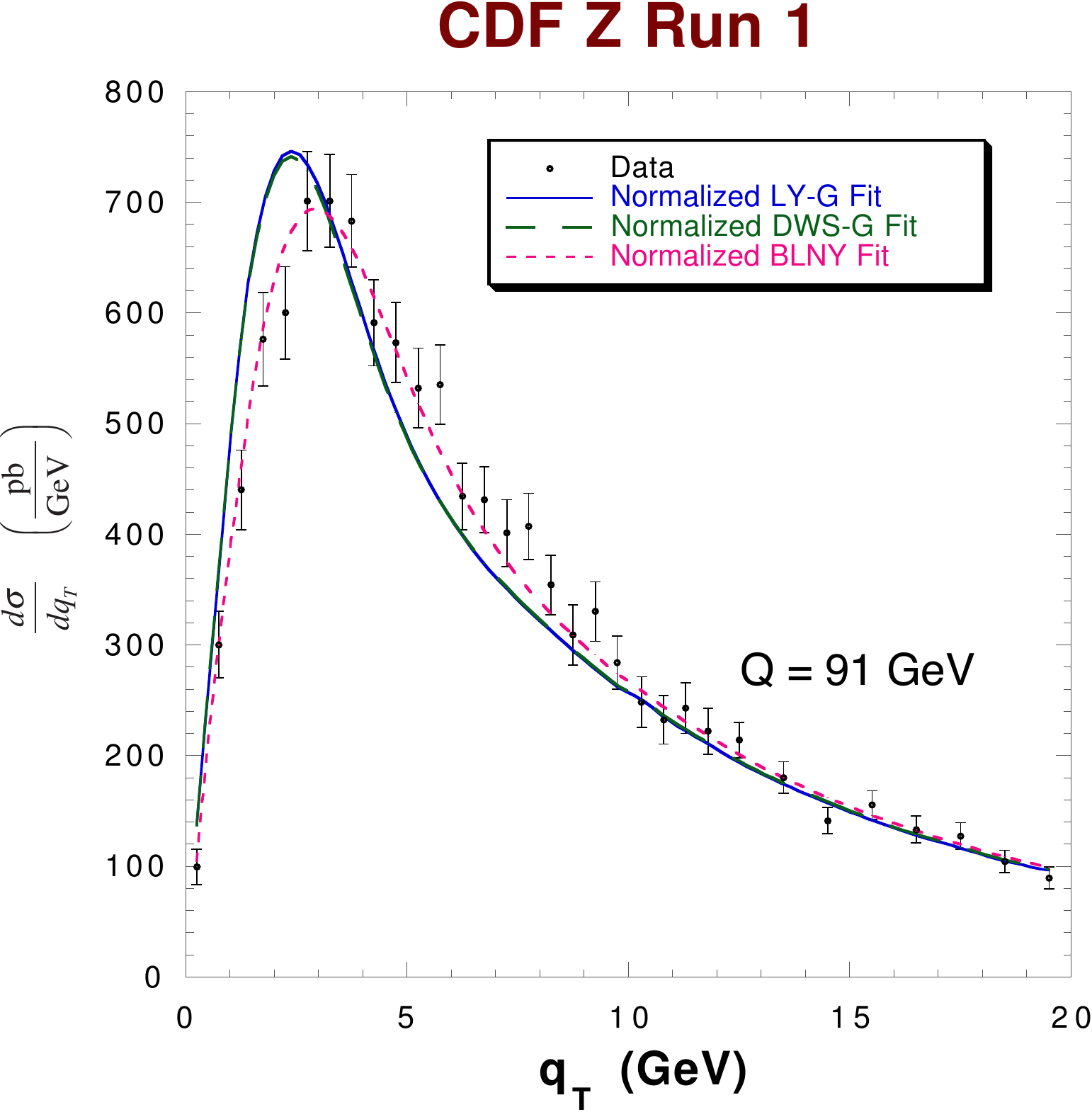}}
  \\
     $\Tsc{b}$
     &
     \VC{\includegraphics[scale=\thisscale]{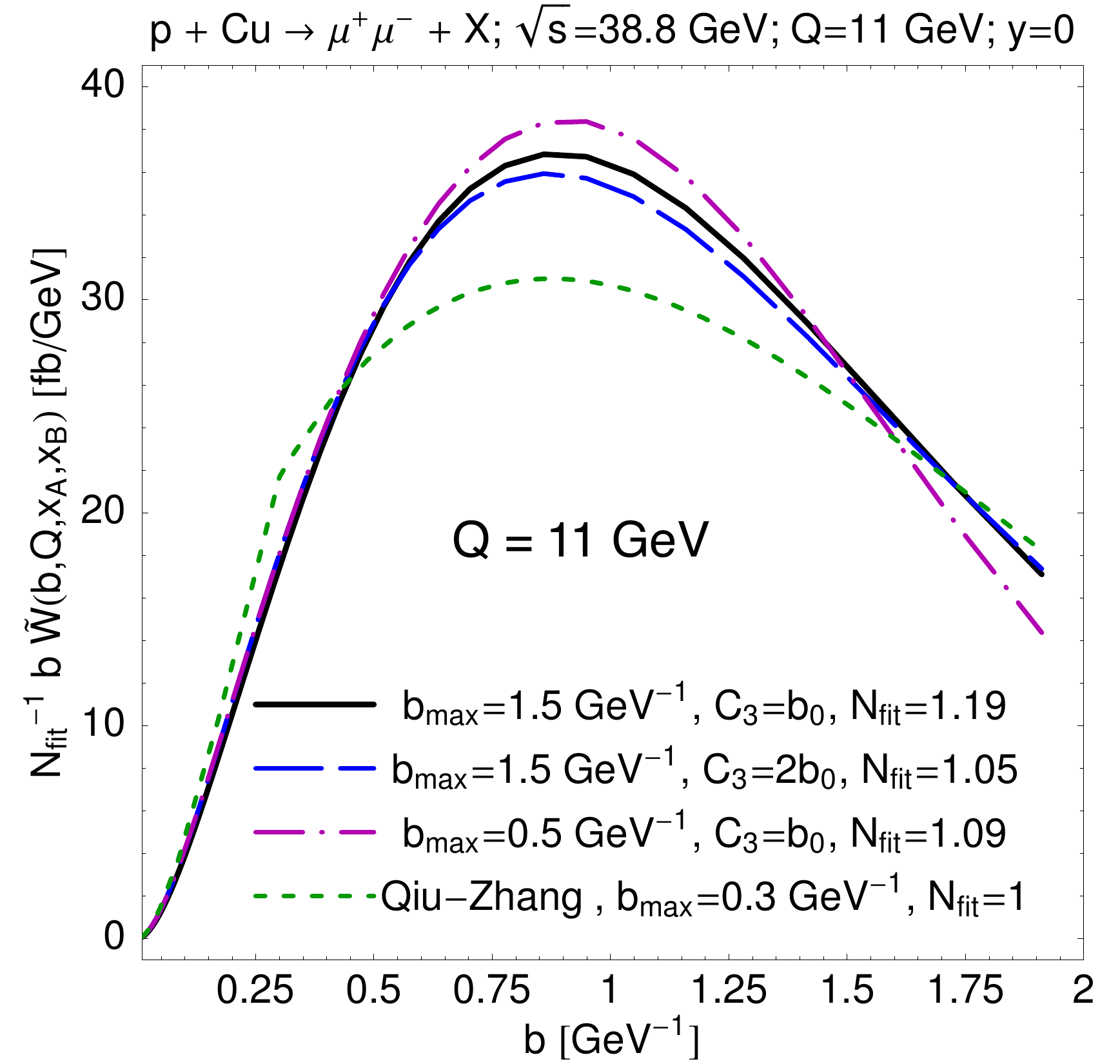}}
     &
     \VC{\includegraphics[scale=\thisscale]{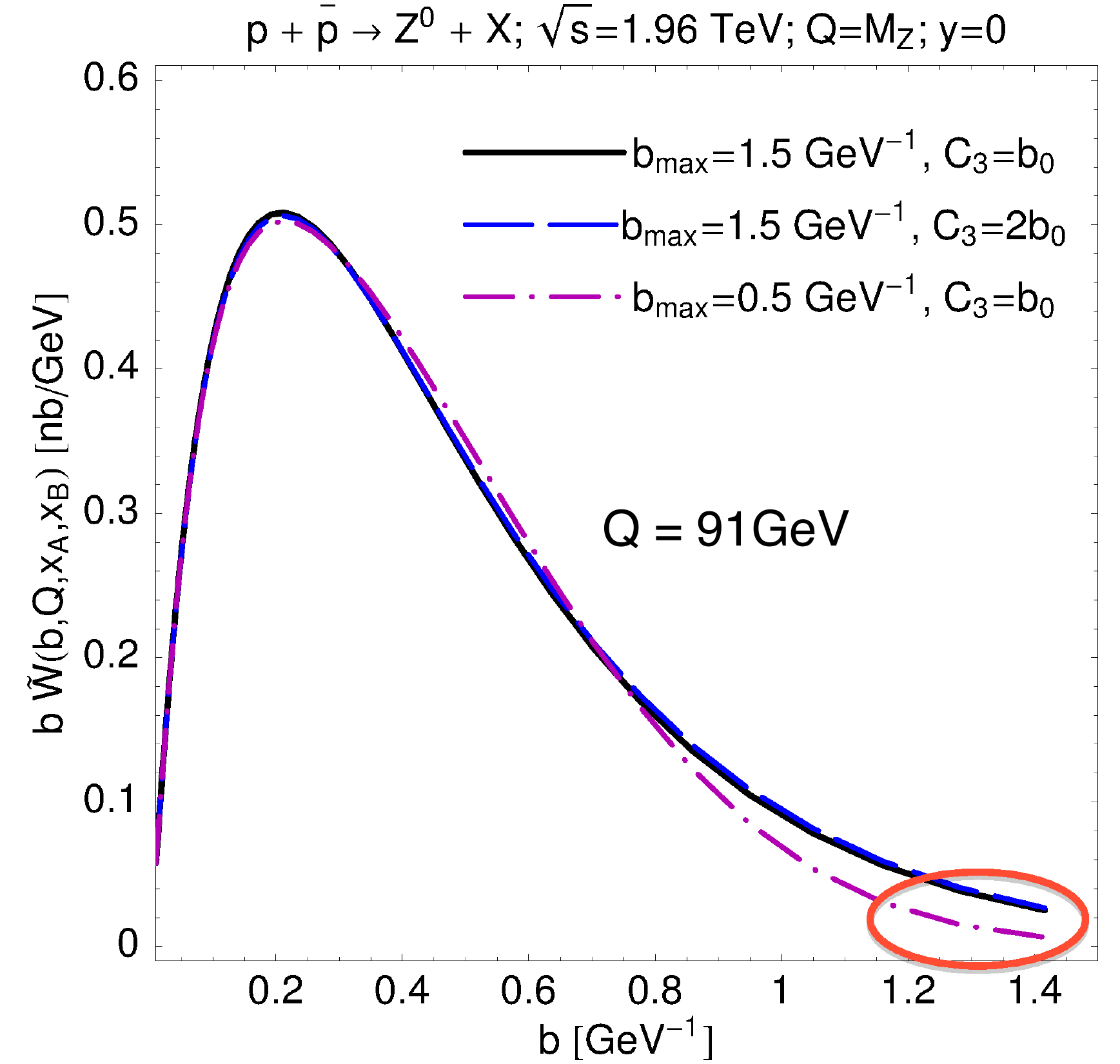}}
  \\
     &
     \small
     $Q$: $\unit[7\mbox{--}18]{GeV}$, $\sqrt{s}=\unit[38.8]{GeV}$ 
     &
     \small
     $Q=m_Z$, $\sqrt{s}=\unit[1800]{GeV}$ 
  \end{tabular}
  \caption{Left, top: cross-section for $Q$ from $\unit[7]{GeV^{-1}}$
    to $\unit[18]{GeV^{-1}}$ in the E605 experiment compared with fits
    by Landry et al.\ (BLNY), adapted from \cite{Landry:2002ix}.
    Left, bottom:
    corresponding $\Tsc{b}$ integrand at $Q=\unit[11]{GeV^{-1}}$  for
    fits by Konychev and Nadolsky (KN) \cite{Konychev:2005iy}.
    Right: similar plots for $Z$ production at the Tevatron; note
    that the cross section is for $\diff{\sigma}/\diff{\Tsc{q}}$
    instead of $\diff{\sigma}/\diff[2]{\T{q}}$, which gives a
    kinematic zero at $\Tsc{q}=0$.
  }
  \label{fig:DY.geography}
\end{figure}

As illustrated in Fig.\ \ref{fig:DY.geography}, evolution to higher
$Q$ shifts the dominant region of $\Tsc{b}$ to ever lower values.
Therefore the differential cross section as a function of $\Tsc{q}$
broadens as $Q$ increases.  The increasing suppression of the large
$\Tsc{b}$ region implies that the cross section is eventually
dominated by perturbative effects, even at $\Tsc{q}=0$, provided
that the large distance properties are non-pathological.

\section{Results for $\tilde{K}$ from fits}

The evolution of the integrand $\tilde{W}$ in Eq.\ (\ref{eq:fact}) is
given by
\begin{equation}
\label{eq:tilde.W.evol}
  \frac{ \diff{ \ln \tilde{W} } }
       { \diff{ \ln Q^2 } }
  =
  G(\alpha_s(Q))
  - \int_{\mu_{\bstarsc}}^Q \frac{\diff{\mu}}{\mu} \gamma_K(\alpha_s(\mu))
  + \tilde{K}(\bstarsc;\mu_{\bstarsc})
  - g_K(\Tsc{b};\bmax) ,
\end{equation}
where $G(\alpha_s(Q))$ is perturbatively calculable.  The right-hand
side is the sum of a $Q$-dependent term and a $\Tsc{b}$-dependent
term. The $Q$-dependent term affects only the normalization of the
cross section.

The change in shape of the cross section is governed only by the
$\Tsc{b}$-dependent part, which can be considered as the value of
$\tilde{K}(\Tsc{b};\mu_1)$ at some fixed reference scale $\mu_1$:
\begin{equation}
\label{eq:tilde.K.mu1}
   \tilde{K}(\Tsc{b};\mu_1)
  =
  - \int_{\mu_{\bstarsc}}^{\mu_1} \frac{\diff{\mu}}{\mu} \gamma_K(\alpha_s(\mu))
  + \tilde{K}(\bstarsc;\mu_{\bstarsc})
  - g_K(\Tsc{b};\bmax) .
\end{equation}
Therefore we can gain an understanding of the change of shape of the
$\Tsc{q}$-dependent cross section with $Q$ from the functional
dependence of $\tilde{K}(\Tsc{b};\mu_1)$ on $\Tsc{b}$.

\begin{figure}
  \centering
    \includegraphics[angle=0,scale=0.6]{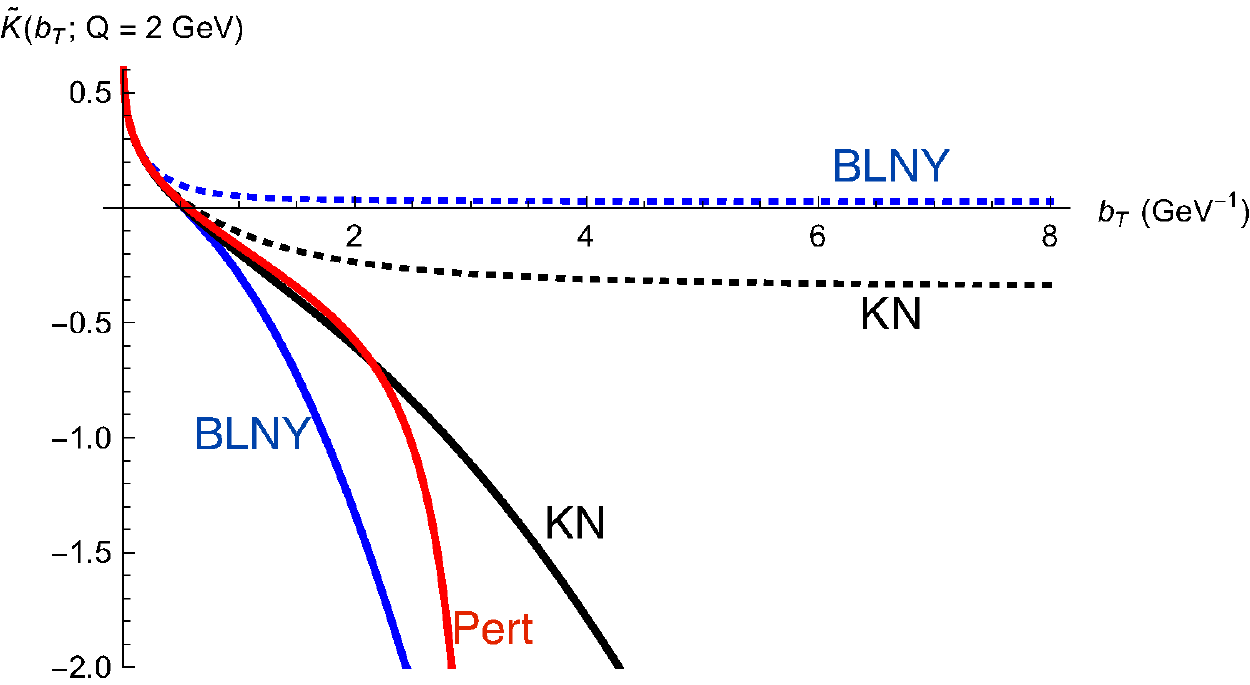}
\\[2mm]
    \begin{tabular}[c]{cccc}
      Typical $\Tsc{b}$: & $\unit[0.5]{GeV^{-1}}$ & $\unit[1.2]{GeV^{-1}}$ & $\unit[3]{GeV^{-1}}$
    \\
      $Q$:                & $m_Z$ &  $\unit[10]{GeV}$ & $\unit[2]{GeV}$
    \end{tabular}
  \caption{$\tilde{K}$ and its components from the KN and BLNY fits.
    See the text for details.
  }
  \label{fig:Ktilde.KN.BLNY}
\end{figure}

In Fig.\ \ref{fig:Ktilde.KN.BLNY} is shown the $\Tsc{b}$ dependence of
$\tilde{K}(\Tsc{b};\unit[2]{GeV^{-1}})$ in the KN and BLNY fits.
Overall, we see a decreasing function, which corresponds to the shift
to smaller dominant values of $\Tsc{b}$ with increasing $Q$ that we
saw in Fig.\ \ref{fig:DY.geography}.  

The red curve gives the purely perturbative,
renormalization-group-improved prediction for $\tilde{K}$.  This is
obtained by setting $\bmax=\infty$ and $g_K=0$ on the right-hand side
of (\ref{eq:tilde.K.mu1}).  The calculation shown was made with 2-loop
approximations for the evolution of the running coupling and for
$\gamma_K$. 
This prediction is valid when $\Tsc{b}$ is small enough, but then
diverges to $-\infty$ from the Landau pole in the approximated
coupling, where perturbative-based calculations are completely
untrustworthy.

CSS's $\bstar$-prescription gives a smooth cut off of the perturbative
term, intended to restrict it to a region where perturbatively based
calculations are valid.  The cut-off part of $\tilde{K}$, i.e., the
first two terms on the right of (\ref{eq:tilde.K.mu1}) is shown as the
dotted curves in Fig.\ \ref{fig:Ktilde.KN.BLNY}.  The blue curve is
for $\bmax=\unit[0.5]{GeV^{-1}}$, corresponding to the BLNY fit of
\cite{Landry:2002ix}.  The black curve is for
$\bmax=\unit[1.5]{GeV^{-1}}$, corresponding to the KN fit
\cite{Konychev:2005iy}, which gives a better fit to the Drell-Yan data.

Finally we include the fitted functions $g_K$ for the two fits giving
the black and blue solid curves. In both cases $g_K$ is purely
quadratic: $g_K\propto \Tsc{b}^2$.  We first notice that each of these
two curves matches the purely perturbative red curve beyond where the
$\bmax$ cut off is important.  Indeed the KN curve gives a good match
to above $\unit[2]{GeV^{-1}}=\unit[0.4]{fm}$, which is, a priori, a
region of at best marginal applicability of perturbative methods. That
the fits match a perturbative calculation suggests that the primary
result of fitting the one parameter in each fit is to reproduce
perturbation theory and then to extrapolate the result to large
$\Tsc{b}$ in a non-singular fashion.

At large $\Tsc{b}$, there is a dramatic difference between the KN and
BLNY curves. This is reflected in the factor of two difference between
the corresponding integrands at the right-hand edge of the lower right
plot in Fig.\ \ref{fig:DY.geography}.  But that large fractional
difference is in a place where the integrands are small, and so has
little effect on the quality of the fits.  

To understand where the differences between the curves matter, we need
to know the typical values of $\Tsc{b}$ involved.
These are given below the
graph in Fig.\ \ref{fig:Ktilde.KN.BLNY}, and we deduce that the fits were
primarily sensitive to $\Tsc{b}$ below about $\unit[2]{GeV^{-1}}$.
The values at larger $\Tsc{b}$ are only an
extrapolation, which need not be correct.  To probe larger values of
$\Tsc{b}$ experimentally, we need lower $Q$.

That the extrapolation is actually wrong is indicated
phenomenologically by the results of Sun and Yuan \cite{Sun:2013hua}.
At large-$\Tsc{b}$, the fitted Gaussian for the TMD pdfs combined with
evolution governed by a quadratic $g_K$ gives a $Q$-dependent Gaussian
behavior for the integrand $\tilde{W}$:
\begin{equation}
  \tilde{W} 
   \sim \dots e^{ - \Tsc{b}^2 [\text{coeff}(x) + \text{const} \ln (Q^2/Q_0^2)] }
   = \dots e^{ - \Tsc{b}^2 a(Q,x)  }
\qquad \mbox{(at large $\Tsc{b}$)}.
\end{equation}
At a value of $Q$ appropriate for data from HERMES and CLAS, the value
of $a(Q,x)$ in this equation becomes negative when the BLNY fit is
used; this gives a completely unphysical cross section. The KN fit
(with $\bmax=\unit[1.5]{GeV^{-1}} = \unit[0.3]{fm}$) merely gives a
value of $a(Q,x)$ much too low to agree with the data.

\section{Improved large-$\Tsc{b}$ properties}

Phenomenologically, we have good standard fits to Drell-Yan data that
determine $\tilde{K}$ for $\Tsc{b}$ up to around $\unit[1.5]{GeV^{-1}}
= \unit[0.3]{fm}$, but not much further.  At lower $Q$, larger
$\Tsc{b}$ dominates, but the extrapolated evolution appears to be too
strong to agree with data. 
Furthermore, from the theoretical side the Gaussian large-$\Tsc{b}$ behavior of
TMD functions is disfavored \cite{Schweitzer:2012hh}. Instead, the
expected behavior of Euclidean correlation functions is an exponential
times a power:
\begin{equation}
   \frac{1}{\Tsc{b}^p} e^{-m\Tsc{b}}.
\end{equation}
That is a non-perturbative statement, with $m$ being the mass of a
relevant state.  (One could, of course, ask whether this expectation
is really correct in a confining theory like QCD)

Supposing that the mass in the exponent is energy-independent suggests
that $\tilde{K}(\Tsc{b})$ goes to a constant as $\Tsc{b}\to\infty$.
The constant gives a $Q$-dependent change (presumably a decrease) to
the normalization of TMD pdfs at large $\Tsc{b}$, but does not affect
the exponential itself.

We therefore proposed \cite{Collins:2014jpa} a new parameterization
for $g_K$:
\begin{equation}
\label{eq:gK.new}
 g_K(\Tsc{b};\bmax)
    = g_0(b_{\rm max}) \left(1 - \exp \left[ - \frac{C_F
          \alpha_s(\mubstar) \Tsc{b}^2}{\pi g_0(b_{\rm max})\,
          \bmax^2} \right] \right) \,,
\end{equation}
where
\begin{equation}
  g_0(\bmax) = g_0({\bmax}_{,0}) + \frac{2 C_F}{\pi}
  \int_{C_1/{\bmax}_{,0}}^{C_1/\bmax} \frac{d \mu^\prime}{\mu^\prime}
    \alpha_s(\mu^\prime)
\end{equation}
is the value of $g_K$ at $\Tsc{b}=\infty$.
This is arranged to have the following properties: 
\begin{itemize}
\item At moderate $\Tsc{b}$ it is approximately quadratic, and the sum
  of $g_K$ and the cut-off $\tilde{K}(\bstarsc;\mu_{\bstarsc})$
  approximately agrees with perturbation theory.
\item At large $\Tsc{b}$, it goes to a constant.
\item The constant is given $\bmax$ dependence to compensate (to
  leading order), the $\bmax$ dependence of the perturbative term:
  $\lim_{\Tsc{b}\to\infty}\tilde{K}(\bstarsc;\mu_{\bstarsc})
  =\tilde{K}(\bmax;\mu_{\bmax})$. 
\item Only one free parameter is used.
\end{itemize}

As we well see, the result gives much reduced $\bmax$ dependence of
evolution compared with standard parameterizations.  The exact
$\tilde{K}$ is a quantity in full QCD and therefore does not itself
have any $\bmax$ dependence whatsoever.  If a simple quadratic
function is used and happens to be correct for $g_K$ at one value of
$\bmax$, it cannot be valid for other values of $\bmax$; a different
functional form is needed.

Of course, we do not imagine that our proposed formula
(\ref{eq:gK.new}) is exactly correct; we simply propose it as a
starting guess from which further refinement is possible.

\begin{figure}
  \centering
  \includegraphics[angle=90,scale=0.4]{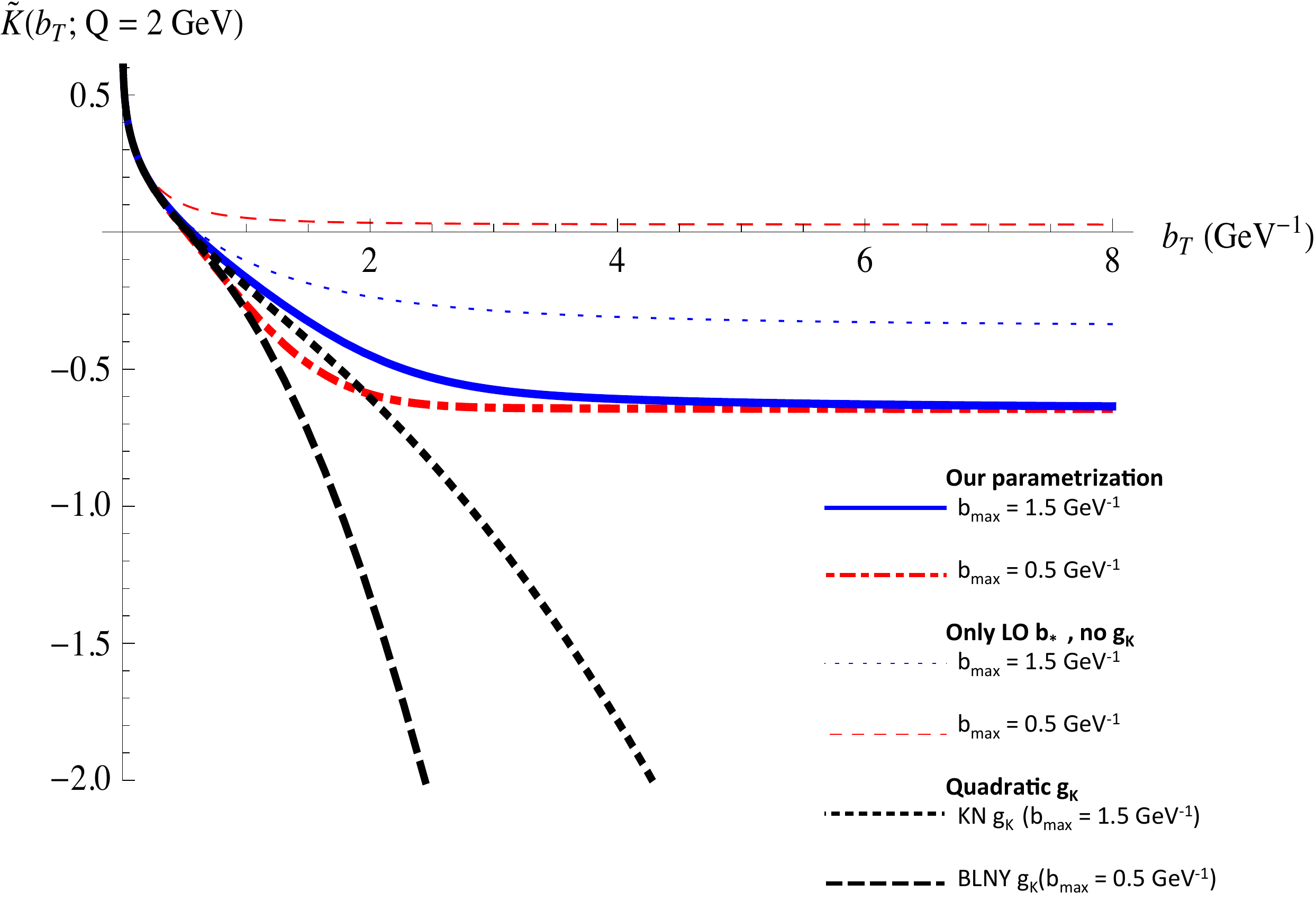}
  \caption{Results of new parameterization for $\tilde{K}$.
   }
  \label{fig:new.K.tilde}
\end{figure}

We have not yet used our new parameterization to perform comparisons
or fits with data.  To show that it is likely to give reasonable
results, we show in Fig.\ \ref{fig:new.K.tilde} what it gives for
$\tilde{K}$ for two values of $\bmax$, with the choice that $g_0=0.3$
when $\bmax=\unit[1.5]{GeV^{-1}}$.  It gives reasonable agreement with
the KN results in the region of $\Tsc{b}$
where the KN fit was determined by Drell-Yan data.  But the flattening of
the curves at larger $\Tsc{b}$ shows that the evolution of the shape
of the distribution is slower at large $Q$, as is necessary to be
compatible with data at lower energy.

\begin{figure}
  \centering
  \includegraphics[scale=0.35, viewport=50 120 750 510, clip]{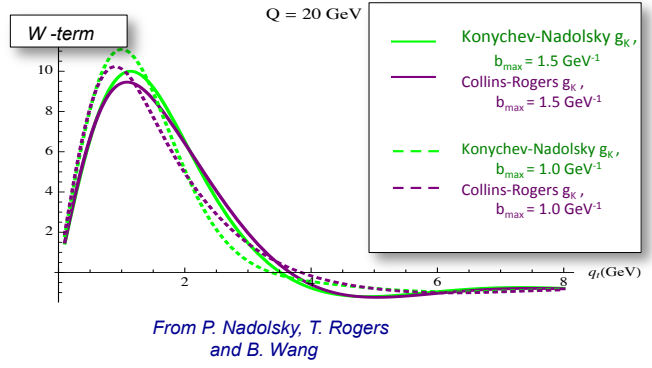}
  \caption{$\tilde{W}$ with new parameterization, compared with KN version.}
  \label{fig:W.new}
\end{figure}

Finally, to measure the compatibility with the Drell-Yan data, Fig.\
\ref{fig:W.new} shows some results (from Nadolsky, Rogers and Wang)
for the Fourier transform in (\ref{eq:fact}), at $Q=\unit[20]{GeV}$,
in comparison with the KN results for two values of $\bmax$.

\section{Summary of results}

We argue that $\tilde{K}$ goes to a constant at large $\Tsc{b}$.  This
value should be measured, of course.  We propose a new
parameterization to interpolate between the constant at large
$\Tsc{b}$ and the known behavior at moderate $\Tsc{b}$.  It should
give better agreement with data and general principles over a wide
range of $Q$.  Further refinement is possible, of course.

To facilitate comparison between different work, fits should be
presented in terms of the full $\tilde{K}(\Tsc{b})$, not just in terms
of $g_K$.

In \cite{Collins:2014jpa}, we argued that the following function can
give useful diagnostics:
\begin{equation}
  A(\Tsc{b})
  =
    - \frac{ \partial }{ \partial\ln \Tsc{b}^2 }
      \frac{ \partial }{ \partial\ln Q^2 }
      \ln \tilde{W}(\Tsc{b},Q,x_A,x_B)
   \stackrel{\textrm{CSS}}{=}
    - \frac{ \partial }{ \partial\ln \Tsc{b}^2 }
      \tilde{K}(\Tsc{b},\mu).
\end{equation}
It controls evolution of shape of TMD functions, is scheme and scale
independent, and is strongly universal.

Acknowledgment: This material is based upon work supported in part by
the U.S. Department of Energy under Grant No.\ DE-SC0008745.


\end{document}